\documentclass[a4paper, 10pt, conference]{ieeeconf}      % Use this line for a4 paper

\IEEEoverridecommandlockouts               
\usepackage{cite}

   \usepackage[pdftex]{graphicx}
\usepackage{amsmath}
\usepackage{algorithmic}
\usepackage{colortbl}
\usepackage{algorithmic}
\usepackage{array}
\usepackage{textcomp}
\usepackage{soul,xcolor}
\usepackage{siunitx}
\usepackage{multirow}
\usepackage{url}
\usepackage{threeparttable}
\usepackage{textcomp}
\usepackage{threeparttable}
\usepackage[ruled,vlined]{algorithm2e}

\def\ps@IEEEtitlepagestyle{%
  \def\@oddfoot{\mycopyrightnotice}%
  \def\@evenfoot{}%
}
\makeatletter
\newcommand*\titleheader[1]{\gdef\@titleheader{#1}}
\AtBeginDocument{%
  \let\st@red@title\@title
  \def\@title{%
    \bgroup\normalfont\large\centering\@titleheader\par\egroup
    \vskip1.5em\st@red@title}
}
\makeatother

\title{Exploiting Nanoelectronic Properties of Memory Chips for Prevention of  IC-Counterfeiting}
\titleheader{Accepted as conference paper at the 22nd IEEE International Conference on Nanotechnology (IEEE-NANO 2022)}
\begin{document}
%\title{\LARGE \bf
%Exploiting Nanoelectronic Properties of Memory Chips for Prevention of  IC-Counterfeiting
%}

\author{Supriya Chakraborty, Tamoghno Das, and Manan Suri 
\thanks{This work was supported in part by  Ministry of Education (MoE), Government of India, SERB under  Grant CRG/2018/001901, PSA office under Grant Prn.SA/Nanoelectronics/2017, and DST under Grant DST/TDT/AMT/2017/159(G).}% <-this % stops a space
\thanks{The authors are with Department of Electrical Engineering, Indian Institute of Technology Delhi, India. (Corresponding author -
        {\tt\small manansuri@ee.iitd.ac.in.})}%
}

\maketitle
%\thispagestyle{empty}
%\pagestyle{empty}

%%%%%%%%%%%%%%%%%%%%%%%%%%%%%%%%%%%%%%%%%%%%%%%%%%%%%%%%%%%%%%%%%%%%%%%%%%%%%%%%
\begin{abstract}

This study presents a methodology for anti-counterfeiting of Non-Volatile Memory (NVM) chips. In particular, we experimentally demonstrate a generalized methodology for detecting (i) Integrated Circuit (IC) origin, (ii) recycled or used NVM chips, and (iii) identification of used locations (addresses) in the chip. Our proposed methodology inspects latency and variability signatures of Commercial-Off-The-Shelf (COTS) NVM chips. The proposed technique requires low-cycle ($\sim$100) pre-conditioning and utilizes Machine Learning (ML) algorithms. We observe different trends in evolution of latency (sector erase or page write) with cycling on different NVM technologies from different vendors. ML assisted approach is utilized for detecting IC manufacturers with 95.1\% accuracy obtained on prepared test dataset consisting of 3 different NVM technologies including 6 different manufacturers (9 types of chips).
\end{abstract}

\footnote{\copyright 2022 IEEE. Personal use of this material is permitted. Permission from IEEE must be obtained for all other uses, in any current or future media, including reprinting/republishing this material for advertising or promotional purposes, creating new collective works, for
resale or redistribution to servers or lists, or reuse of any copyrighted component of this work in other works.}

%%%%%%%%%%%%%%%%%%%%%%%%%%%%%%%%%%%%%%%%%%%%%%%%%%%%%%%%%%%%%%%%%%%%%%%%%%%%%%%%
\vspace{-1mm}
\section{INTRODUCTION}

Counterfeiting of Integrated Circuits (ICs) is a significant concern for the semiconductor supply chain \cite{guin2014counterfeit}. Reliability and security issues of counterfeited or degraded Non-Volatile Memory (NVM) chips have the potential to threaten various sectors such as automotive, defense/security, medical devices, consumer electronics, etc. \cite{guin2014counterfeit,kumari2018independent}. Variation in nanoscale properties of different NVM technologies occurs due to variation in underlying nanostructures and nanomaterials. Various invasive and non-invasive techniques exist in literature which can be employed on Commercial-Off-The-Shelf (COTS) chips to detect IC origin \cite{talukder2020towards} or recycled ICs \cite{guin2014counterfeit,kumari2018independent,guin2014low,guo2017ffd}.  In addition, other security primitives like chip identity information, memory-based Physical Unclonable Functions (PUFs) \cite{kim2015investigation}, True Random Number Generator (TRNG) \cite{wang2012flash}, etc. are implemented to track NVM chips. However, most of these techniques are proposed for NAND Flash memory \cite{kumari2018independent} or DRAM \cite{talukder2020towards} chips. Moreover, detailed chip characterization and maintenance of large database for ascertaining COTS NVM chip authenticity is cumbersome.  Techniques like physical inspection, statistical analysis \cite{kumari2018independent}, introduction of extra circuitry \cite{guin2014low}, etc. for counterfeit IC identification are either hefty or ineffective when involving many variants of NVM chips, or require modification in CMOS circuitry during fabrication. Significant advancement in Machine Learning (ML) techniques facilitates employing ML algorithms for counterfeit IC detection \cite{aramoon2020impacts}.  In this study, we present a non-invasive anti-counterfeiting technique based on exploiting the nanoelectronic phenomena (switching time variability) of COTS NVM chips. In particular, we exploit latency (for sector erase/page write operation) to provide a solution for (i) identification of the IC manufacturer, and (ii) recycled IC detection for conventional and emerging NVM technologies. The key contributions of this paper are:
\begin{enumerate}
    \item A generalized methodology to identify (a) NVM chip manufacturer, (b) recycled NVM chips, and (c) used locations on the chip. 
    \item The identification methodology exploits the switching time variability of NVM chips and requires $\sim$100 program/erase cycles to be performed on the chip.
    \item Generation of custom dataset to train an ML based classifier that can effectively determine IC origin/manufacturer.
\end{enumerate}
The rest of the paper is organized as follows: Section II provides an outline of NVM technologies used. Experimental results
and discussion are presented in Section III. Section IV concludes the paper.

\begin{figure}[!t]
\centering
\includegraphics[scale=0.29]{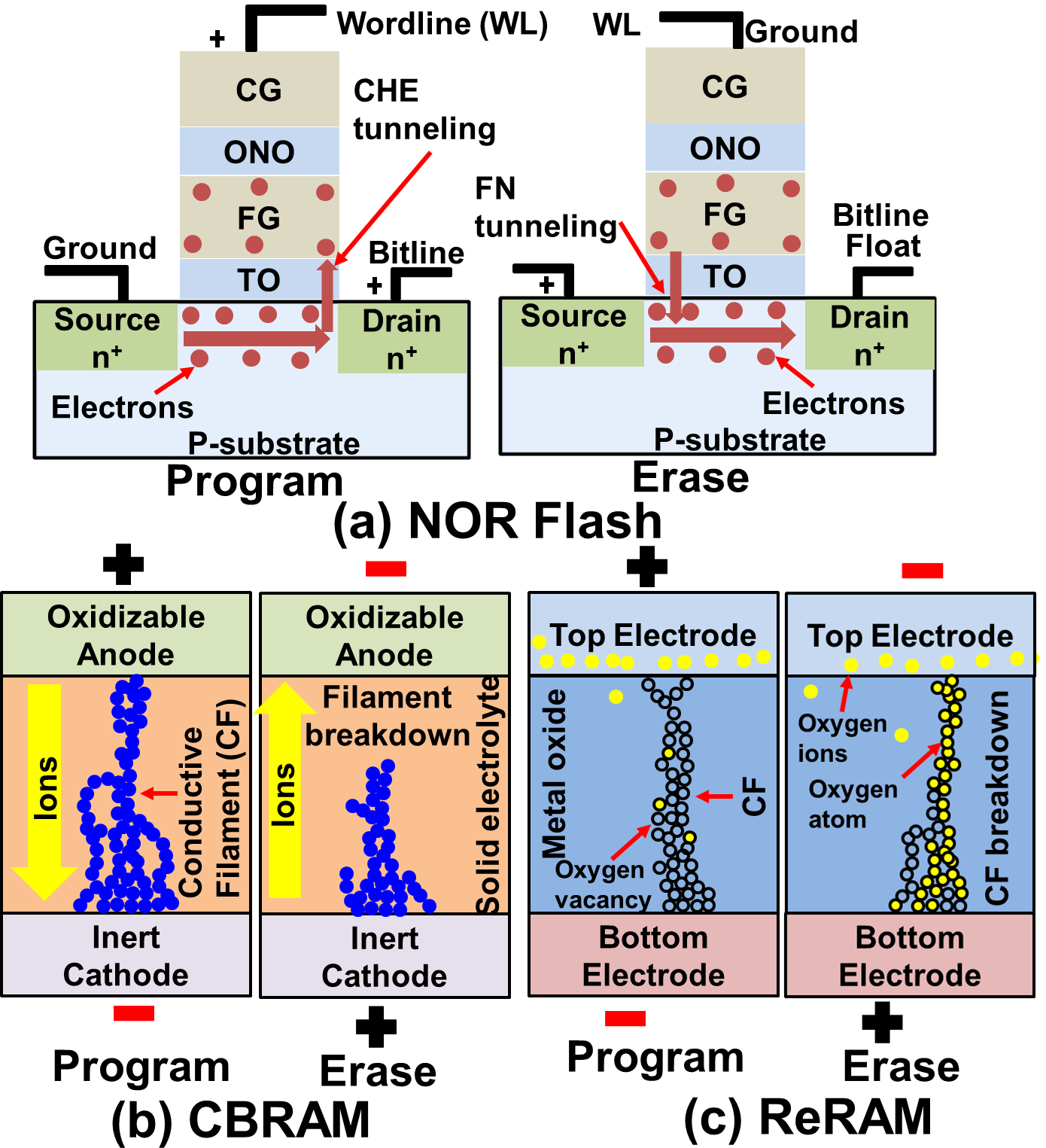}
\caption{Working principle of different NVM technologies. For CBRAM and RRAM, program = set state (LRS), and erase = reset state (HRS). }
\label{device}
\vspace{-7mm}
\end{figure}
\begin{figure}[!t]
\centering
\includegraphics[scale=0.32]{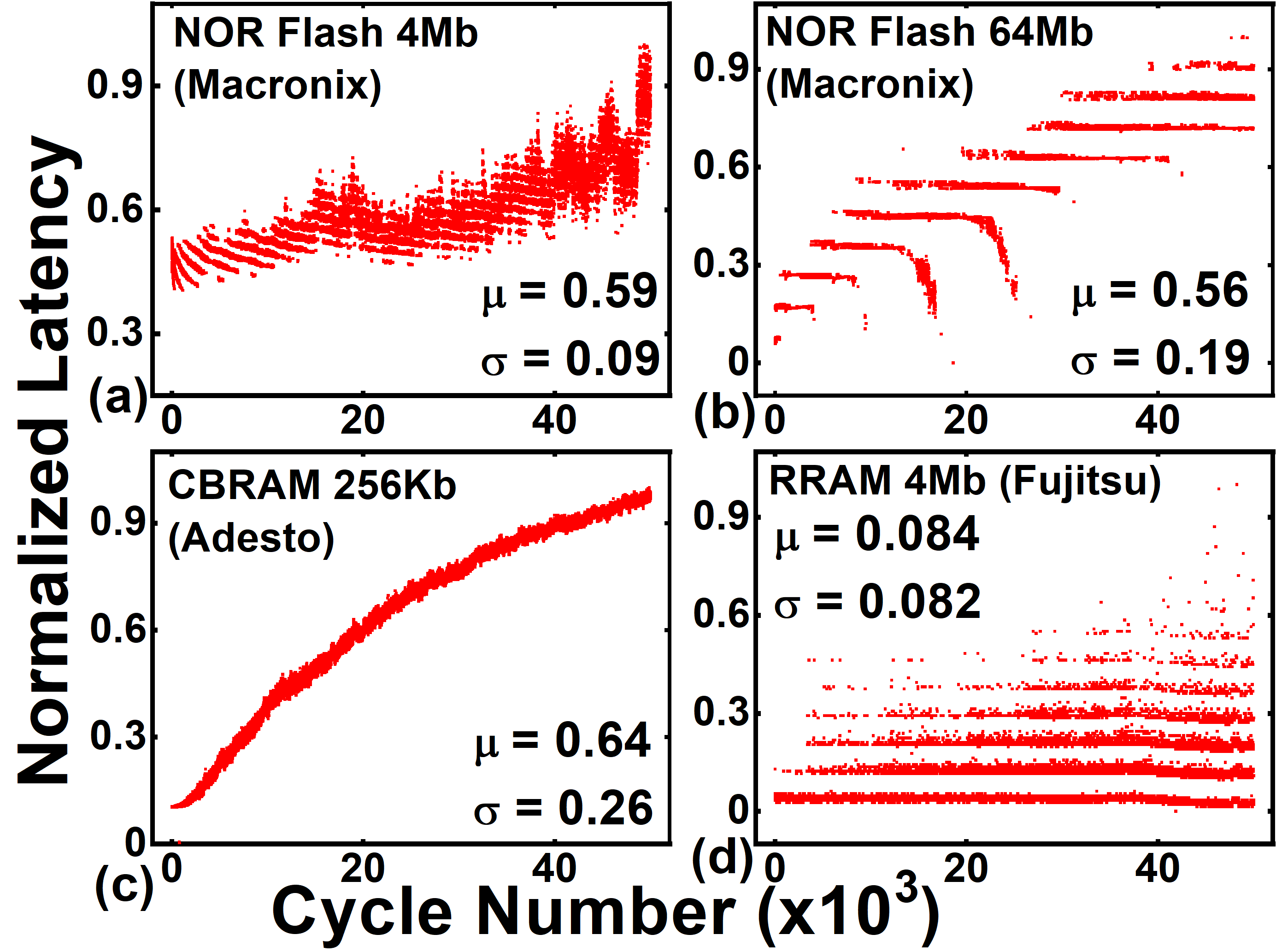}
\caption{(a)-(d) Evolution of sector erase/page write latencies over cycling (50k cycles) for various NVM chips. Distinct evolution in latency values is observed for different NVM technologies and manufacturers.}
\label{latency_all}
\vspace{-3mm}
\end{figure}

\begin{figure}[!t]
\centering
\includegraphics[scale=0.25]{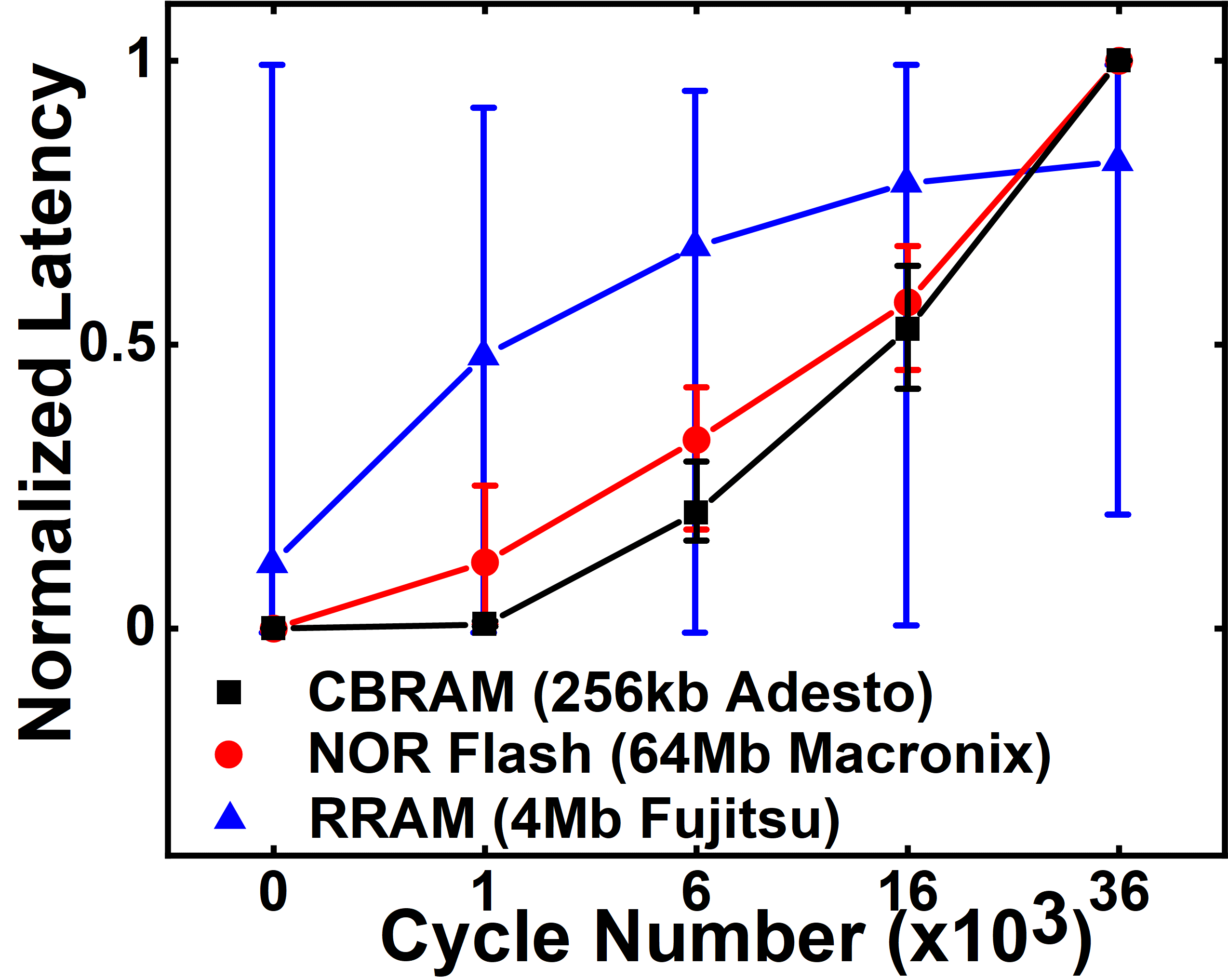} 
\caption{Statistics of latency values obtained within a span of 50
cycles before and after 1k, 6k, 16k and 36k cycles. Each point
= 2 chips with 5 random locations in each chip (1k samples).}
\label{stat_lat}
\vspace{-7.5mm}
\end{figure}

\vspace{-1mm}
\section{OUTLINE OF NVM TECHNOLOGIES USED}
NOR Flash memory technology utilizes Metal-Oxide-Semiconductor (MOS) structure with a stacked-double-poly Floating-Gate (FG). The gate is bounded by a dielectric material and electrically controlled by a capacitive-coupled-Control-Gate (CG). The working principle of NOR Flash memory is based on threshold voltage (Vth) modulation of FG transistor. During program operation [WordLine (WL) and BitLine (BL) = positive, source = ground], electrons are trapped into the FG through the dielectric by Channel-Hot Electron (CHE) injection mechanism. During erase operation (WL = ground, BL = float, source = positive), electrons are detrapped from the FG following the Fowler–Nordheim (FN) tunneling mechanism. This trapping (high Vth) and detrapping (low Vth) results in changing of Vth of the FG transistor (high Vth = logic “1,” and low Vth = logic “0”) [shown in Fig. \ref{device}(a)] \cite{bez2003introduction}. A particular class of emerging Resistive Random Access Memory (RRAM) technology with metal-insulator-metal device structure, Conductive-Bridging Random Access Memory (CBRAM), works on the formation and dissolution of Conductive Filament (CF). Applying voltage of specific polarity across the device terminals results in formation of CF as a consequence of diffusion and reduction of ions/defects. Applying voltage of reverse polarity (for bipolar devices) leads to dissolution of the CF [Fig. \ref{device}(b)]. Formation and dissolution of the CF switch the devices to Low-Resistance State (LRS/ON) and High-Resistance State (HRS/OFF), respectively. The state of the device (HRS/LRS) exhibits the bit stored in the memory cell \cite{kozicki2016conductive}. Another class of RRAM technology consists of metallic layers or stacks as top and bottom electrodes and a metal oxide insulating layer interposed between the electrodes (metal-oxide-metal structure). The working of these RRAM technologies is based on (i) migration of oxygen ions (ii) thermal dissolution. Initially, the fresh devices are formed due to soft dielectric breakdown and drifting of oxygen ions towards the anode. CF is formed in the bulk oxide due to either oxygen vacancies or metal precipitates \cite{wong2012metal}. During the programming operation (LRS) current flows through the CF to the electrodes. During erase operation (HRS), oxygen ions migrate back to the bulk leading to filament breakdown due to recombination [as shown in Fig. \ref{device}(c)]\cite{wong2012metal}.

\vspace{-1mm}
\section{EXPERIMENTAL RESULTS AND DISCUSSION}
\subsection{Experimental Setup}
The test setup consists of a soft-core processor implemented on an FPGA evaluation board. The processor performs program and erase operations on the NVM chip. The NVM chip is connected to FPGA board through PMOD connectors. 
Detailed description of the experimental setup is provided in \cite{chakraborty2020true}. Extensive experiments are performed on different SPI based NVM chips i) NOR Flash, ii) CBRAM, and iii) RRAM from multiple vendors (listed in Table \ref{nvm_list}) at room temperature with identical operating conditions. Latency measurements for sector erase operation in NOR Flash and page write operation in CBRAM and RRAM chips are performed by monitoring the Write-In-Progress (WIP) bit \cite{chakraborty2020true} in the status register. Erase latency is measured by performing continuous 4kB sector erase operations. Page write latency is measured by initially programming a page with a known data value, followed by programming it with a new data value having bit-flips at particular locations. The latencies of NVM chips are measured successfully as the switching time variations of NVM devices (within ms or \SI{}{\micro s}) used for our study are within the range of the time resolution of our setup (10 ns).  

\begin{table}[!t]
\renewcommand{\arraystretch}{1}
\caption{Details of NVM chips used}
\label{nvm_list}
\centering
\begin{tabular}{|c|m{1.8cm}|m{1.8cm}|c|}
\hline
Manufacturer&Capacity (class tag)  & IC Part No.& Technology\\
\hline
Macronix&4Mb (class0), 64Mb (class1), and 128Mb (class2) &MX25R4035F, MX25L6445E, MX25L12835F&NOR Flash \\
\hline
Micron&128Mb (class3)&RM25C256DS&NOR Flash\\
\hline
Winbond& 8Mb (class4)&W25Q80DV&NOR Flash \\
\hline
Microchip &2Mb (class5)&SST25PF020B&NOR Flash\\
%\hline
%Adesto&16 Mb&AT25SF161&NOR Flash\\
\hline
Adesto&256kb (class6)&RM25C256DS&CBRAM\\
\hline
Fujitsu&4Mb (class7), 8Mb (class8)&MB85AS4MT, MB85AS8MT&RRAM\\
\hline
 \end{tabular}
\vspace{-6mm}
\end{table}

\vspace{-2mm}
\subsection{Electrical Characterization Results and Analysis}
Fig. \ref{latency_all}(a)-(d) shows the evolution of measured latency values for different NVM chips with cycling (50k cycles).
Difference in evolution pattern of the obtained latency values is observed within different NVM technologies and manufacturers. We observe unique latency signatures on comparing different chips of same NVM technology spread across different memory capacities and manufacturers. We also compare the latency signatures among different NVM technologies. The possible reasons for uniqueness in signature may be: variation in chip architecture, technology node, layout design, material stack, process (PVT) variations, etc. \cite{guin2014counterfeit,talukder2020towards}. The distinctiveness in latency evolution patterns can thus be used as a fingerprint to predict the NVM chip origin (manufacturer and capacity). Fig. \ref{stat_lat} shows the statistics of latency values for different NVM technologies. The obtained data includes spatial and chip-to-chip variability. We observe significant increase in latency values with cycling compared to the initial values. Continuous programming/erase operations on NVM chip lead to device degradation. The nanoscale causes of NVM device degradation over cycling include:- increase in number of defects due to high electric field operation, increase in read and write noise, change in distribution of threshold voltage, random migration of ions, aging of devices, trapping of charges, change in distribution of shape, height, thickness, radius of conductive filaments, degradation of vacancy mobility, generation of extra vacancy, etc. \cite{kumari2018independent,liu2018optimization,841251,arita2015switching,chen2011physical,huang2013analytic}. All these factors in combination affect the reliability of chip performance. The device degradation in NVM chip may cause slowing down of operation (erase/write) \cite{kumari2018independent}. Subsequently, a higher number of programming pulses may be required for performing reliable operations. This leads to increase in effective latency of the used sectors/pages \cite{sakib2018non,huang2013analytic}. Thus, change in erase/program latency over cycling is a good metric to differentiate between used and unused memory chips.
%(statistics shown in Fig. \ref{framework}).  

\begin{table}[!t]
\renewcommand{\arraystretch}{1}
\caption{Performance comparison of ML algorithms for IC prediction}
\label{ML_comp}
%\centering
\begin{tabular}{|m{1.6cm}|m{1.8cm}|m{1.1cm}|m{1cm}|m{1cm}|}
\hline
Feature selection technique& ML algorithm &  Test accuracy (\%) & Inference time (s) & Training time (s)\\ 
%\cline{3-8}
\hline
\multirow{1}{*}{No feature }  &Decision Tree & 86.1 &0.0545 & 389.39\\
\cline{2-5}
selection (All & KNN & 95.1& 0.1957 & 748.43\\
\cline{2-5}
100 features) & Gaussian SVM & 90.1 & 0.72 & 971.31 \\
\hline
\multirow{1}{*}{MRMR test } &Decision Tree & 87.9& 0.0346 & 89.54\\
\cline{2-5}
  (25 features)& KNN & 91.7& 0.09 & 120.11\\
\cline{2-5}
 & Gaussian SVM &89.0 & 0.1125 & 332.69\\
\hline
\multirow{1}{*}{NCA  } &Decision Tree &  89.2&0.034 & 98.74\\
\cline{2-5}
  (25 features)& KNN & 92.4& 0.0783 & 123.66\\
\cline{2-5}
 & Gaussian SVM &88.3 & 0.1286 & 315.93\\
\hline
\end{tabular}
\vspace{-6mm}
\end{table}

\vspace{-2mm}
\subsection{Proposed Methodology of Detecting IC Manufacturer}
%\subsection{Dataset and ML Algorithms}
We investigate different ML algorithms for predicting IC origin (manufacturer and capacity). We generate the test dataset by performing repeated erase/program operations at random locations in a fresh chip. This is followed by sampling the latency in continuous groups of 100 values at distinct intervals (after 0, 1k, 5k, 10k, 15k, 30k, and 50k cycles) to capture various stages of usage of the chips and their spatial variability. Each of the measured 100 latency values in a group is used as separate feature for training and testing.
%Since erase/program latency is an effective metric in distinguishing used and unused memory chips as discussed in Section III-B.
It is experimentally observed that the effect of cross-correlation among classes on classification accuracy is overcome when using 100 consecutive latency values as features.
The dataset spans 9 chip types (9 classes) from 6 manufacturers, encompassing 3 different NVM technologies (specified in Table \ref{nvm_list}). Comprehensive experiments are performed on 3 chips of each memory type to include chip-to-chip variability. The total dataset constitutes $\sim$2250 sample points of which 80\% of instances are used for training and 20\% of instances are used for testing. We define the output class designating manufacturer with particular memory capacity. 
\begin{figure}[!t]
\centering
\includegraphics[scale=0.32]{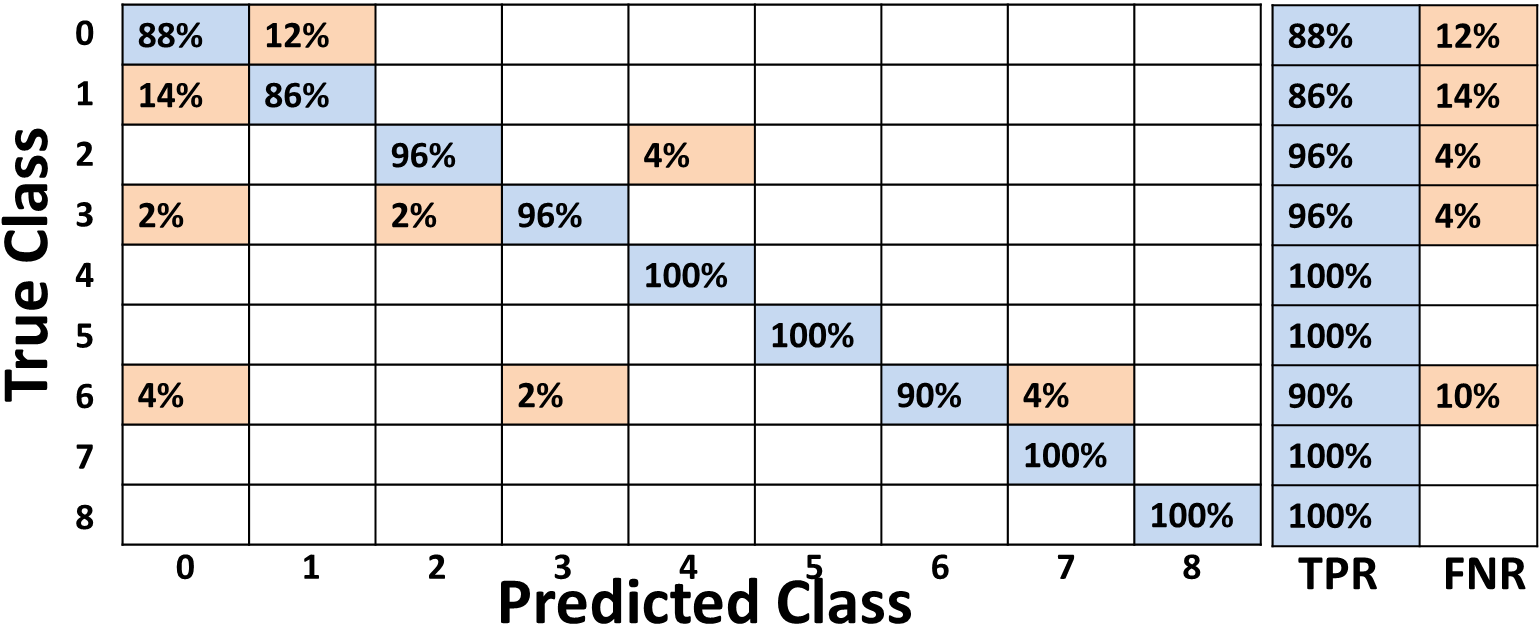} 
\caption{Test confusion matrix for KNN classifier. TPR = True Positive Rates, FNR = False Negative Rates.}
\label{confusion}
\vspace{-7mm}
\end{figure}

For the implementation of ML algorithms, we investigate different feature selection techniques like Minimum Redundancy Maximum Relevance (MRMR) algorithm \cite{peng2005feature} and Neighbourhood Component Analysis (NCA)  \cite{goldberger2004neighbourhood}, which are commonly used for optimal feature selection in classification problems. Training of all three classifiers (Decision Tree,  K-Nearest Neighbors (KNN), and Gaussian Support Vector Machine (SVM)) is performed either on all 100 features or on a subset of features (obtained using aforementioned feature selection techniques) by 8-fold cross-validation technique. Table \ref{ML_comp} shows the comparative performance of the three different ML algorithms, including two different feature selection techniques. It is observed that the KNN classifier trained using all 100 features, shows the best accuracy (95.1\%) on test data. Additionally, use of feature selection techniques drastically reduces the time required for training and inference, whilst maintaining reasonably high accuracy. Fig. \ref{confusion} reports the test accuracy for KNN classifier obtained for each particular type of chip (technology-wise accuracy). We observe that all types of chips can be predicted quite accurately. We utilize our ML classifier to build a swift IC-counterfeiting detection methodology that can predict the actual manufacturer of a given NVM chip by performing $\sim$100 program/erase cycles at any location on the chip, and measuring the latency values in each cycle. These values are then passed to our ML classifier for predicting the actual chip type. 

\begin{figure}[!t]
    \centering 
\includegraphics[scale=0.30]{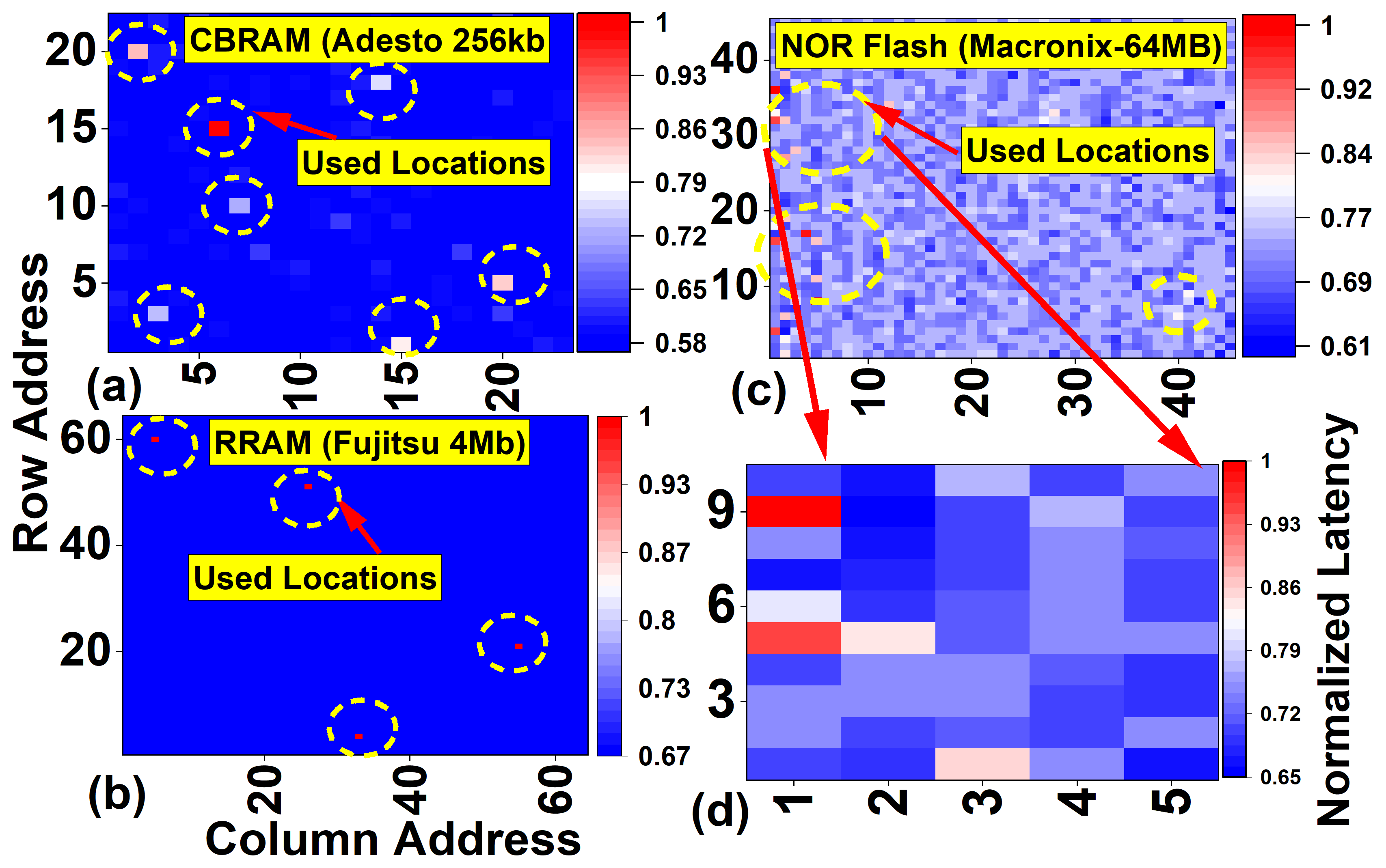}
\caption{Spatial latency map of full chip including all addresses for (a) CBRAM, (b) RRAM and (c) NOR Flash. Random locations are selected in a fresh chip and are programmed/erased continuously to generate used locations artificially. Dotted circles show such used locations in the NVM chips. (d) Zoomed view of used and unused individual locations for a NOR Flash chip.}
\label{lat_spatial}
\vspace{-5mm}
\end{figure}

Further, we extend the technique to identify used locations in the NVM chip. To experimentally validate the concept, we generate used locations artificially by carrying out 1k, 5k, 10k, 20k, 30k, and 50k program/erase cycles at random locations. We measure individual latency values (in between the specified program/erase cycles) for all sectors/pages in an entire chip by performing a single program/erase operation on all the locations of the chip. Fig. \ref{lat_spatial}(a)-(c) shows the obtained spatial latency map of entire chip over all addresses individually for different NVM technologies. We observe a relative increase in latency of used (erased/programmed) locations (as explained in section III-B) compared to unused sectors/pages in the chip by a factor of $\sim$1.5X - 2.5X. Table \ref{comparison} summarizes the proposed methodology and presents a comparative analysis with existing literature on memory IC anti-counterfeiting techniques. The proposed technique offers a generalized solution against IC-counterfeiting for conventional and emerging NVM chips. In contrast to the previous works, we have used 3 different technologies including 6 vendors. The proposed methodology demonstrates reasonable accuracy in ML based IC anti-counterfeiting for multiple classes of NVM chips. In future extensions of this work, we would like to extend the proposed methodology to predict the range of usage (number of program/erase cycles performed) of the used locations. This can be achieved using the magnitudes of obtained latency values of the predicted chip type. 
We would also like to explore additional features like current consumption and advanced pattern recognition techniques based on deep neural networks.

 \begin{table}[!t]
\renewcommand{\arraystretch}{1}
\caption{Comparison of memory IC anti-counterfeiting techniques}
\label{comparison}
\centering
\begin{tabular}{|m{1.4cm}|m{1.3cm}|m{1.3cm}|m{1.3cm}|m{1.3cm}|}
\hline
Ref. No.& \cite{kumari2018independent} & \cite{talukder2020towards} &\cite{sakib2021flash}&\textbf{This work*}\\
\hline
Parameter Observed&Latency and error &Latency and error&Partial erase and error & Program and erase latency \\
\hline
Technology&NAND Flash (MLC)&DRAM&NAND Flash (SLC)& NOR Flash (SLC), CBRAM, RRAM\\
\hline
Cell granularity&Block&Page&Page&Sector/page\\
\hline
ML technique&-& One class classifier&-& K-NN\\
\hline
Manufacturer Variants&1&3&4&6\\
\hline
Accuracy(\%)&100&-&-&95.1\\
%\hline
%Minimum usage detection&$\sim$150 PE cycles&-&-&1k-15k cycles\\
\hline
 \end{tabular}
\vspace{-7mm}
\end{table}

\vspace{-1mm}
\section{CONCLUSION}
We present an anti-counterfeiting technique for detecting IC origin (capacity and manufacturer) and recycled (or used) chips, along with used locations. The proposed methodology exploits intrinsic property variations within NVM COTS chips. We experimentally illustrate that latency and variability can be used to detect counterfeit NVM chips. Finally, we present ML based approach for detecting authentic IC manufacturers with high accuracy. A dedicated dataset through experimental characterization is also developed for conventional and emerging NVM technologies to train the ML classifier to high accuracy.

%\vspace{-1mm}
\bibliographystyle{IEEEtran}
\bibliography{IEEEabrv,bibliography.bib}

\end{document}